\documentclass[aps,twocolumn,nofootinbib]{revtex4-2}

\usepackage{graphicx}
\usepackage{amssymb, amsmath,mathtools}
\usepackage{bbm}
\usepackage{enumerate}
\usepackage{subfigure}
\usepackage[colorlinks=true]{hyperref}
\usepackage{orcidlink}

\graphicspath{{./}{./plots/}}

\usepackage{xcolor}

\allowdisplaybreaks

\begin{document}

\title{Spontaneous breaking of the $\text{SO}(2N)$ symmetry in the Gross-Neveu model}

\author{SangEun Han\,\orcidlink{0000-0003-3141-1964}}
\affiliation{Department of Physics, Simon Fraser University, Burnaby, British Columbia, Canada V5A 1S6}

\author{Igor F.~Herbut\,\orcidlink{0000-0001-5496-8330}}
\affiliation{Department of Physics, Simon Fraser University, Burnaby, British Columbia, Canada V5A 1S6}

\begin{abstract}

The canonical Gross-Neveu model for $N$ two-component Dirac fermions in $2+1$ dimensions suffers a continuous phase transition at a critical interaction $g_{c1} \sim 1/N$ at large $N$, at which its continuous symmetry $\text{SO}(2N)$ is preserved and a discrete (Ising) symmetry becomes spontaneously broken. A recent mean-field calculation, however, points to an additional transition at a different critical $g_{c2}\sim -N g_{c1}$, at which $\text{SO}(2N) \rightarrow \text{SO}(N) \times \text{SO}(N)$. To study the latter phase transition we rewrite the Gross-Neveu interaction $g (\bar{\psi} \psi)^2$ in terms of three different quartic terms for the single ($L=1$) $4N$-component real (Majorana) fermion, and then extend the theory to $L>1$. This allows us to track the evolution of the fixed points of the renormalization group transformation starting from $L\gg 1$, where one can discern three distinct critical points which correspond to continuous phase transitions into (1) $\text{SO}(2N)$-singlet mass-order-parameter, (2) $\text{SO}(2N)$-symmetric-tensor mass-order-parameters, and (3) $\text{SO}(2N)$-adjoint nematic-order-parameters, down to $L=1$ value that is relevant to the standard Gross-Neveu model. Below the critical value of $L_c (N)\approx  0.35 N$ for $N\gg1$ only the Gross-Neveu critical point (1) still implies a diverging susceptibility for its corresponding  ($\text{SO}(2N)$-singlet) order parameter, whereas the two new  critical points that existed at large $L$ ultimately become equivalent to the Gaussian fixed point at $L=1$. We interpret this metamorphosis of the $\text{SO}(2N)$-symmetric-tensor fixed point from critical to spurious as an indication that the transition at $g_{c2}$ in the original Gross-Neveu model is turned first-order by fluctuations.
\end{abstract}

\maketitle

Gross-Neveu model \cite{gross, rosenstein, zinnjustinbook} in 2+1 dimensions provides probably the simplest example of fermionic criticality, and as such it has been well studied over the years. Besides its methodological importance in high-energy physics  its variants have also been connected to quantum phase transitions in condensed matter systems that involve gapless quasi-relativistic fermions such as graphene \cite{herbutprl, hjr}, unconventional superconductors \cite{vojta, huh},  and surfaces of topological insulators \cite{herbutfisherbook}. In these systems the leading instabilities at strong couplings are typically toward Lorentz-invariant order-parameters that represent relativistic mass terms for low-energy Dirac fermions, and which translate into broken-symmetry insulating or superconducting states in terms of the original electrons.

It was recently shown that all but one such different mass-order-parameters for $N$ copies of two-component Dirac fermions can be unified into a single  representation of the $\text{SO}(2N)$ symmetry of the free Dirac Lagrangian in 2+1 dimensions \cite{so8}. One mass-order-parameter is always a singlet, whereas the remaining ones transform as a symmetric irreducible two-component tensor under $\text{SO}(2N)$. The singlet, in the context of graphene where $N=4$ for example, corresponds to quantum anomalous Hall state \cite{haldane} which breaks a discrete Ising (time-reversal) symmetry.  The remaining 15 insulating and 20 superconducting mass-order-parameters \cite{qed32, mudry, herbut2007, herbut2012} fall into the symmetric tensor representation of dimension 35 of $\text{SO}(8)$. Furthermore, the mean-field calculation suggests that if the interaction term broke the $\text{SO}(8)$ in favor of an insulating ground state at half-filling, a finite chemical potential would eventually cause a first-order flop into a superconductor \cite{so8}.

 The main technical novelty in Ref.~\cite{so8} is the set of Fierz identities \cite{hjr} that enable rewriting of the standard Gross-Neveu interaction term $g (\bar{\psi} \psi)^2$ as a sum of squares of the fermion bilinears that form the symmetric tensor representation of $\text{SO}(2N)$. Since this exact transformation also involves a change of the overall sign of the interaction term, it reveals that the tensor and the singlet mass-order-parameters are in direct competition, with the winner at strong coupling depending on the sign. The  standard  mean-field calculation shows that whereas for one sign of the Gross-Neveu interaction $g$, $\text{SO}(2N)$ symmetry is preserved and the Ising symmetry becomes broken at a critical value $g_{c1}$ through the usual Gross-Neveu transition, for the opposite sign of $g$ there exists a new critical value, $g_{c2}$, beyond which $\text{SO}(2N)$ symmetry becomes spontaneously broken to $\text{SO}(N)\times \text{SO}(N)$, and some of the components of the tensor order parameter develop an expectation value. Both transitions at the mean-field level appear continuous, and for large $N$ at least, $g_{c2}\sim (-N) g_{c1}$. The $\text{SO}(2N)$-preserving Gross-Neveu transition, however, can also be understood beyond the mean-field theory, where, for example, it is found to remain continuous. Its critical exponents have been computed by $1/N$ and $\epsilon$-expansions, and recently by conformal bootstrap \cite{vasilev, gracey1, gracey2, zinnjustin, erramilli}. Such a deeper understanding of the second transition at $g_{c2}$ at present is lacking. In this paper we make the first step in this direction by formulating a renormalization group approach to the $\text{SO}(2N)$-symmetry breaking transition of the Gross-Neveu model.

The Gross-Neveu model in $2+1$ Euclidean dimensions is defined by the action $S=\int d\tau d^2 x (\mathcal{L}_0 + \mathcal{L}_1)$, and
\begin{align}
\mathcal{L}_0={}& \psi^\dagger (\mathbb{I}_{N} \otimes (\mathbb{I}_{2} \partial_\tau -i \sigma_1 \partial_1 - i \sigma_3 \partial_2 ))   \psi,\\
\mathcal{L}_1 ={}&  g (\psi^\dagger (\mathbb{I}_{N} \otimes \sigma_2)   \psi)^2,
\end{align}
where $\tau$ is the imaginary time, $\mathbb{I}_{N}$ is the $N$-dimensional unit matrix, and $\sigma_{i}$ ($i=1,2,3$) is the Pauli matrix. The cutoff $\Lambda$ on momentum  and frequency integration may be assumed, and we set the velocity of fermions to unity. $\psi$ is the $2N$-component (complex) Dirac fermion. The Gross-Neveu model is invariant under continuous $\text{U}(1)$, $\text{SU}(N)$  and Lorentz symmetries, as well as under the discrete time-reversal, as well known \cite{so8}.  Note that we chose all Pauli matrices featured in the  kinetic energy part $\mathcal{L}_0$ to be symmetric, with the antisymmetric $\sigma_2$ appearing only in the interaction term $\mathcal{L}_1$. This allows one to straightforwardly introduce the ``real", or ``Majorana" fermions as $\psi = (\phi_1 - i \phi_2)/\sqrt{2}$ and $\psi^\dagger = (\phi_1 ^{\intercal} + i \phi_2^{\intercal})/\sqrt{2} $, in terms of which the kinetic and the interacting parts of the Lagrangian may be rewritten as
\begin{align}
\mathcal{L}_0={}& \frac{1}{2}  \phi^{\intercal}  (\mathbb{I}_{2N} \otimes (\mathbb{I}_{2} \partial_\tau -i \sigma_1 \partial_1 - i \sigma_3 \partial_2 ))   \phi ,\\
\mathcal{L}_1 ={}& \frac{g}{4} (\phi^{\intercal} (\mathbb{I}_{2N} \otimes \sigma_2) \phi)^2,\label{eq:symm_int}
\end{align}
where $\phi^{\intercal} = (\phi_1 ^{\intercal}, \phi_2 ^{\intercal} ) $ is the $4N$-component {\it real} fermion. In this ``Majorana representation" it becomes evident that, besides being Lorentz invariant,  the Gross-Neveu Lagrangian in 2+1 dimensions is also invariant under the transformation $\phi \rightarrow (O \otimes \mathbb{I}_{2}) \phi$, where the orthogonal real matrix $O \in \text{SO}(2N)$, and the enlarged group $\text{SO}(2N) \supset \text{U}(1)\times \text{SU}(N)$ \cite{so8}.

\begin{figure}[!t]
\centering
\subfigure[]{\includegraphics[width=0.485\linewidth]{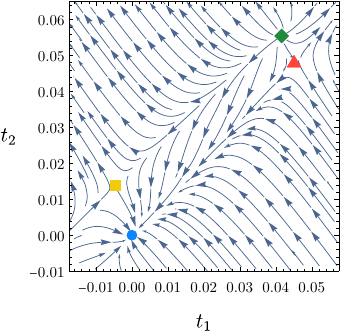}} \hfill
\subfigure[]{\includegraphics[width=0.485\linewidth]{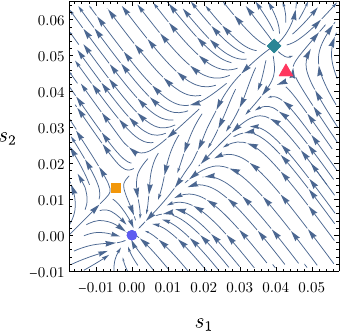}}
    \caption{The RG flow diagrams for the planes (a) $g_{1}+g_{2}+g_{3}=-1/(4(N-1))$ and (b) $g_{1}+g_{2}+g_{3}=0$ when $L=1$ and $N=20$. (a) The blue circle, yellow square, green diamond, and red triangle stand for the standard Gross-Neveu fixed point, bicritical points 1, 2, and unstable fixed point, respectively. Here, $g_{1}=-1/(4(N-1))-(t_{1}-t_{2})/\sqrt{2}$, $g_{2}=t_{1}/\sqrt{2}$, and $g_{3}=-t_{2}/\sqrt{2}$. (b) The indigo circle, orange square, teal diamond, and pink triangle stand for the Gaussian, $\text{SO}(2N)$ symmetric, nematic fixed points, and bicritical points, respectively. Here, $g_{1}=-(s_{1}-s_{2})/\sqrt{2}$, $g_{2}=s_{1}/\sqrt{2}$, and $g_{3}=-s_{2}/\sqrt{2}$.}\label{fig:RG_flow_L1}
\end{figure}

The identity proven in Ref.~\cite{so8} in terms of the real fermions acquires a particularly compact form:
 \begin{equation}
 (\phi^{\intercal} (\mathbb{I}_{2N} \otimes \sigma_2) \phi)^2 = \frac{-1}{N+1} (\phi^{\intercal} (\mathbb{S}_{b} \otimes \sigma_2) \phi) (\phi^{\intercal} (\mathbb{S}_{b} \otimes \sigma_2) \phi),
 \end{equation}
where the summation is over the repeated index $b=1,2,\cdots,(N+1) (2 N-1)$, which enumerates linearly independent $2N$-dimensional, traceless, symmetric matrices $\mathbb{S}_{b}$, normalized as $\text{Tr} (\mathbb{S}_{a} \mathbb{S}_{b}) = 2 N \delta_{ab}$. Note the overall minus sign on the right-hand side. For $g<g_{c1} \sim (-1) /(N\Lambda)$ the mean-field theory suggests that $\langle \phi^{\intercal} (\mathbb{I}_{2N} \otimes \sigma_2) \phi \rangle  \neq 0$, which is the usual Gross-Neveu transition into the $\text{SO}(2N)$-singlet. Using the Hubbard-Stratonovich (Hartree) decoupling of the right-hand side of the last equation \cite{so8} when $g>0$, however, equally suggests that for $g>g_{c2} \sim (N+1)/(N \Lambda)$  $\text{SO}(2N)$ becomes broken, and some of the components of the symmetric tensor order parameter $\langle \phi^{\intercal} (\mathbb{S}_{b} \otimes \sigma_2) \phi \rangle$ become finite. It is evident, however, that while the (negative) critical value $g_{c1}$ becomes small at large $N$, which facilitates the usual perturbative large-$N$ approach to the standard Gross-Neveu transition, the (positive) critical value $g_{c2}$ does not. One may rightfully question therefore whether the continuous mean-field transition at $g_{c2}$ indeed corresponds to some fixed point of the renormalization group transformation, similarly to $g_{c1}$ albeit at strong coupling, or it does not. In the latter case one may suspect that the quantum fluctuations make the $\text{SO}(2N)$-symmetry-breaking transition at $g_{c2}$ discontinuous.

To address this issue we first notice that the Fierz formulas derived in Ref.~\cite{so8} also imply an additional identity,  in terms of real fermions:
\begin{equation}
(\phi^{\intercal} (\mathbb{I}_{2N} \otimes \sigma_2) \phi)^2 = \frac{-1}{3N} (\phi^{\intercal} (\mathbb{A}_{a} \otimes \sigma_2 \sigma_i) \phi) (\phi^{\intercal} (\mathbb{A}_{a} \otimes \sigma_2 \sigma_i) \phi),
\end{equation}
where the summation on the right-hand side now goes over $i=1,2,3$, and $a= 1,2,\cdots,N(2N-1)$. The latter index counts linearly independent, antisymmetric $2N$-dimensional matrices $\mathbb{A}_{a}$, also normalized as $\text{Tr} (\mathbb{A}_{a} \mathbb{A}_{b}) = 2 N \delta_{a b}$. The fermion bilinear $ \phi^{\intercal} (\mathbb{A}_{a} \otimes \sigma_2 \sigma_i) \phi $ transforms therefore as the adjoint irrep.~of $\text{SO}(2N)$, and as a vector under Lorentz transformation. A condensation of such a bilinear for $i=1$ or $i=3$, for example, would correspond to breaking of the spatial rotational symmetry, i.e.~to a general ``nematic" state \cite{sachdev, schwab}. In graphene, the adjoint irrep.~of $\text{SO}(8)$ would be 28-dimensional, for example. The three interaction terms written as squares of the singlet, symmetric, and antisymmetric $\text{SO}(2N)$ tensor components are therefore all precisely proportional to each other. This conforms to the general theorem \cite{fierz} which allows not more than one linearly independent $\text{SO}(2N)$-symmetric contact quartic term for fermions.

\begin{figure*}
\centering
\subfigure[]{\includegraphics[width=0.235\linewidth]{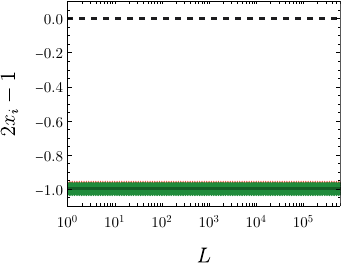}\label{fig:suscep_exp_a}}\hfill
\subfigure[]{\includegraphics[width=0.235\linewidth]{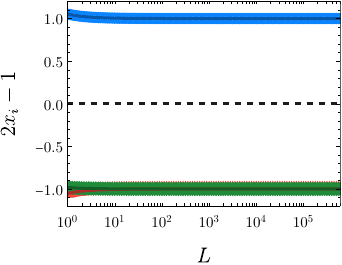}\label{fig:suscep_exp_b}}\hfill
\subfigure[]{\includegraphics[width=0.235\linewidth]{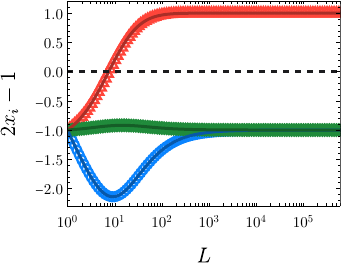}\label{fig:suscep_exp_c}}\hfill
\subfigure[]{\includegraphics[width=0.235\linewidth]{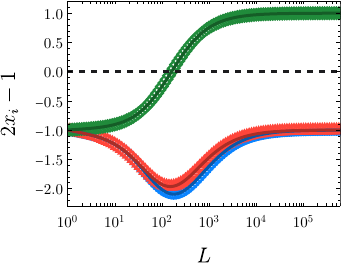}\label{fig:suscep_exp_d}}\\
\subfigure[]{\includegraphics[width=0.235\linewidth]{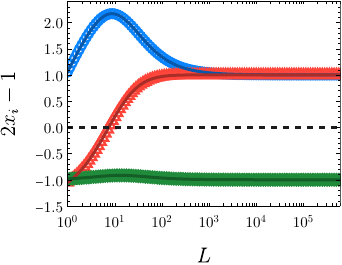}\label{fig:suscep_exp_e}}\hfill
\subfigure[]{\includegraphics[width=0.235\linewidth]{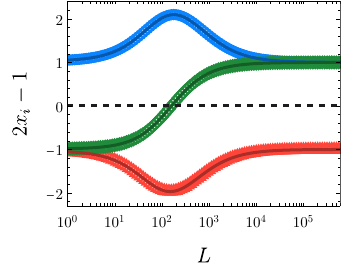}\label{fig:suscep_exp_f}}\hfill
\subfigure[]{\includegraphics[width=0.235\linewidth]{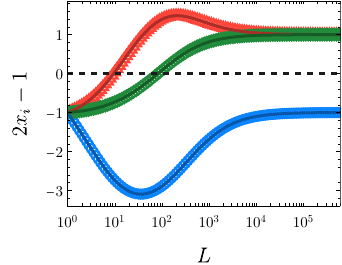}\label{fig:suscep_exp_g}}\hfill
\subfigure[]{\includegraphics[width=0.235\linewidth]{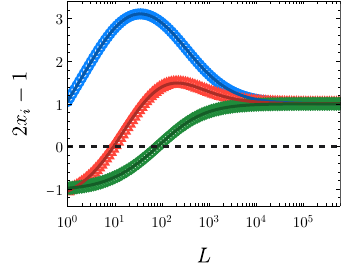}\label{fig:suscep_exp_h}}
\caption{The values of $2x_{i}-1= \gamma_i/\nu$ ($i=1,2,3$) at (a) the Gaussian, (b) Gross-Neveu, (c) $\text{SO}(2N)$ symmetric, (d) nematic, (e-g) bicritical 1-3, and (h) unstable fixed points for $N=20$ and $1\leq L$. The blue circle, red triangle, and green diamond markers stand for $2x_{1,2,3}-1$, respectively. The gray dashed line represents $2x_{i}-1=0$. The corresponding susceptibility diverges when $2x_{i}-1>0$, since the correlation length exponent $\nu>0$ always at a critical point.}\label{fig:suscep_exp}
\end{figure*}

We now generalize the Gross-Neveu model as
\begin{equation}
\mathcal{L}_0 \rightarrow  \frac{1}{2} \phi^{\intercal} _k (\mathbb{I}_{2N} \otimes (\mathbb{I}_{2} \partial_\tau -i \sigma_1 \partial_1 - i \sigma_3 \partial_2 ))   \phi_k ,
\end{equation}
\begin{widetext}
\begin{equation}
\mathcal{L}_1 \rightarrow \frac{g_1}{4} (\phi^{\intercal} _k \mathbb{I}_{2N} \otimes \sigma_2 \phi_k )^2
-\frac{g_2}{4(N+1)} (\phi^{\intercal}  _k  \mathbb{S}_{b} \otimes \sigma_2 \phi_k )^2
-\frac{g_3}{12N} (\phi^{\intercal} _k  \mathbb{A}_{a} \otimes \sigma_2 \sigma_i \phi_k )^2 ,\label{eq:g1g2g3_int}
\end{equation}
 \end{widetext}
 by introducing $L$ flavors of the $4N$-component real fermions, enumerated by the new index $k=1,2,\cdots,L$. When $L=1$ all three terms in Eq.~\eqref{eq:g1g2g3_int} are the same and our representation is {\it redundant}; the interaction term $\mathcal{L}_1$ can simply be rewritten as in Eq.~\eqref{eq:symm_int} with $g=g_1 + g_2 + g_3 $. For $L>1$, however, the three terms in Eq.~\eqref{eq:g1g2g3_int} are readily seen to be linearly independent, and the theory has the larger $\text{SO}(L)\times \text{SO}(2N)$ symmetry \cite{comment}.

 We now perform Wilson's momentum-shell transformation to the leading order in the coupling constants \cite{herbutbook}. Integrating out the real fermions with the magnitude of their three-momenta between $\Lambda/e^l$ and $\Lambda$, the coupling constants are found to run as
 \begin{widetext}
 \begin{align}
 \frac{d g_1}{dl} ={}& -g_1 - 4 (LN -1) g_1 ^2 - 4 (2N-1) (g_2 + g_3) g_1 + \frac{8 (2N-1)}{9 N} g_3 ^2,\label{eq:g1_RG}\\
 \frac{d g_2}{dl} ={}& -g_2 + \frac{4 ( N (L -1) +1)}{N+1}  g_2 ^2 + 4  (g_1 + g_3) g_2 - \frac{8 N}{3} g_2 g_3 -  \frac{8 (N^2 -1)}{9 N} g_3 ^2,\label{eq:g2_RG}\\
 \frac{d g_3}{dl} ={}& -g_3 - \frac{4 (L +1)}{9 }  g_3 ^2 + 4  (g_2 + g_1) g_3 - \frac{ 4N^2}{N+1 } g_2 ^2  -  \frac{28 (N -1)}{9 } g_3 ^2
  - \frac{16 N }{3} g_2 g_3,\label{eq:g3_RG}
 \end{align}
\end{widetext}
where we also have redefined them as $g_i \Lambda /(2 \pi^2 )\rightarrow g_i$ for convenience.

First, we observe that there always exists the ``Gross-Neveu" critical fixed point (``1") of the above transformation at $g_1 ^* = -1/(4 (L N-1) )$, $g_2^* = g_3^* =0$, since the coupling $g_1$ does not generate any of the other two couplings, and the renormalization group transformation is closed with $g_1$ alone. In contrast, both $g_1$ and $g_2$ become generated by $g_3$, and $g_3$ in turn becomes generated by $g_2$; with either $g_2$ or $g_3$ present one needs all three couplings to have a closed set under renormalization. For $L \gg 1$, the equations decouple, and two additional critical points (that is,  fixed points with exactly one infrared relevant direction) become clearly discernable: (2) $g_1^* =O(1/L^4)$,  $g_3^* =O(1/L^2)$, $g_2 ^* = (N+1) / (4 L N )$, and (3) $g_1^* = g_2 ^* =O(1/L^2) $, $g_3 ^* = -9 /(4 L)$. The mean-field analysis for the singlet, symmetric tensor, and nematic order parameters separately finds them finite precisely for $g_1 <  -1/(4 L N )$, $g_2 >  (N+1) / (4 L N )$, and $g_3 < -9 /(4 L)$, respectively, so we identify the critical fixed points (1), (2), and (3) at large $L$ with the mean-field transitions in these three distinct available channels for condensation, respectively. Besides  (1) ``Gross-Neveu," (2) ``symmetric," and (3) ``nematic" critical points, at large $L$ there are also three bicritical, one tricritical, and one completely stable (Gaussian) fixed point of the renormalization group transformation.

\begin{figure}
    \centering
    \includegraphics{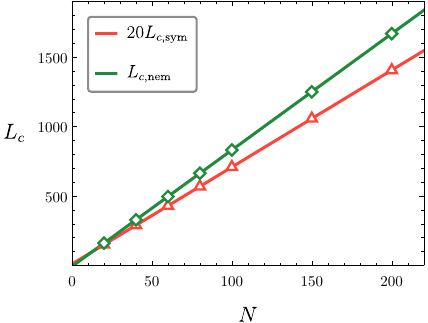}
    \caption{The critical values of $L$ for the susceptibility exponent for given $N$. The solid red and green lines are the fitting functions $L_{c,\text{sym}} = 0.4902 + 0.3490 N$ and $L_{c,\text{nem}} = -7.1975 + 8.3784 N$, respectively. The red diamond and green triangle are the actual critical values of $L$ for given $N$. Note that the values for the symmetric order parameter are magnified 20 times to facilitate comparison.}
    \label{fig:critical_L}
\end{figure}

To substantiate the identification of the critical points in terms of the corresponding order parameters one may compute the susceptibilities for each order parameter at each fixed point: introduce a source-term in the Lagrangian by adding one of the three terms
\begin{align}
\mathcal{L}_{\text{GN}}={}& h_1 (\Phi^{\intercal} _k (\mathbb{I}_{2N} \otimes \sigma_2) \Phi_k) ,\\
\mathcal{L}_{\text{sym}} ={}& h_2 (\Phi^{\intercal} _k (\mathbb{S}_{a} \otimes \sigma_2) \Phi_k) ,\\
\mathcal{L}_{\text{nem}} ={}& h_3 (\Phi^{\intercal} _k (\mathbb{A}_{b} \otimes \sigma_2 \sigma_i) \Phi_k) .
\end{align}
With the change of the cutoff the sources $h_i$ renormalize as \cite{vafek, andras}
\begin{equation}
\frac{d h_i}{dl} = h_i (1+ x_i (g_1, g_2, g_3)).
\end{equation}

The susceptibility for each order parameter near the critical point behaves as
\begin{equation}
\chi_i \sim |\delta g| ^{-\gamma_i },
\end{equation}
where $\delta g$ is the appropriate tuning parameter for the transition, and the exponent $\gamma_i = (2 x_i (g_1 ^*, g_2 ^*, g_3 ^*)-1) \nu$, with $\nu$ as the correlation length exponent at the critical point in question \cite{herbutbook}.  The susceptibility is thus diverging if the corresponding exponent  $\gamma_i$ is positive. Since at the critical points the exponent $\nu$ is always positive, this requires that  $x_i (g_1 ^*, g_2 ^*, g_3 ^*) >1/2$. To the leading order in coupling constants, we find
\begin{align}
x_1 ={}& - 2 (2 N -1) (g_1 + g_2 + g_3)-4N (L-1) g_1,\label{eq:x1}\\
\notag\\
x_2 ={}& 2 (g_1 + g_2 + g_3) + \frac{4N(L-1)}{N+1} g_2,\label{eq:x2}\\
x_3 ={}& -\frac{2}{3} (g_1 + g_2 + g_3) - \frac{4(L-1)}{9}g_3.\label{eq:x3}
\end{align}
At each of the three critical points at large $L$ there is therefore precisely and only one diverging susceptibility, which corresponds to the already identified associated order parameter, and with the corresponding exponent having the mean-field value $\gamma_i=1$. The correlation length exponent at all three critical points at large $L$ is also $\nu=1$.

\begin{figure}
    \centering
    \includegraphics{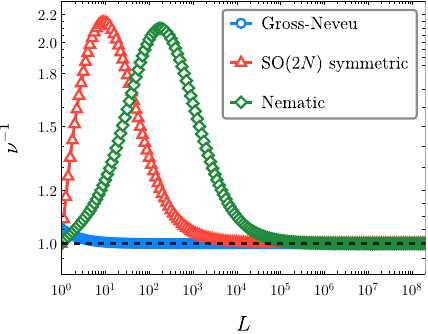}
    \caption{The values of the correlation length exponent $\nu^{-1}$ at the Gross-Neveu (blue circle), $\text{SO}(2N)$ symmetric (red triangle), and nematic (green diamond) fixed points when $N=20$. The dashed gray line stands for $\nu^{-1} =1$. In the large-$L$ limit the three values converge to unity, and match the mean-field value.}
    \label{fig:correl_length_exp}
\end{figure}

It is now interesting to see what happens to the three critical points and the corresponding diverging susceptibilities as the number of Majorana fermion copies $L$  is decreased to the value $L=1$, which corresponds to the original Gross-Neveu model.  First of all, it is clear from Eqs.~\eqref{eq:x1}-\eqref{eq:x3} that all the susceptibilities  depend only on the sum of the three couplings when $L=1$. Furthermore, summing the three flow equations \eqref{eq:g1_RG}-\eqref{eq:g3_RG} one finds
\begin{widetext}
\begin{equation}
\frac{d g}{d l} = -g - 4 (N-1) g^2 -4 N (L-1) ( g_1 ^2 -\frac{ g_2 ^2 }{N+1} + \frac{g_3 ^2 }{9N} ),
\end{equation}
\end{widetext}
where $g=g_1 + g_2 + g_3$. So when $L=1$, all the fixed points are located either in the attractive plane $g^*=0$ (called ``Gaussian plane"), or in the repulsive $ g^* = - 1/(4 (N-1))$ plane (``Gross-Neveu plane"). The flows within these two planes is given as an illustration in Fig.~\ref{fig:RG_flow_L1}, for $N=20$. We find all eight fixed points to survive the limit $L\rightarrow 1$, when $N > 14.2252$. At $N=14.2252$ and $L=1$, two pairs of fixed points, one pair in the Gaussian and another in Gross-Neveu plane, annihilate. Since the one-loop beta-function can be understood as the leading approximation in the large-$N$ limit, hereafter we limit the discussion to $N>14.2252$. Since for $N>14.2252$ there are no collisions of the fixed points as $L$ is decreased all the fixed points retain their character \cite{kaplan, qed31, gukov}, but the numerical values of the susceptibility exponents change. For all $N>14.2252$ we  find the symmetric and the nematic fixed points to end up in the $g=0$ plane, so that although they remain critical in the standard renormalization group sense, all three susceptibility exponents become $\gamma_i =-1$, yielding no divergent susceptibility. Only the Gross-Neveu critical point, which is constrained to be on the line $g_2=g_3=0 $ at any $L$, finds itself in the $g^* = -1/(4(N-1)) $ plane, and therefore keeps its susceptibility for the scalar order parameter divergent. The evolution of the three exponents $\gamma_i/\nu$ at the Gross-Neveu, symmetric, nematic, and the remaining five fixed points with $L$ is presented at Fig.~\ref{fig:suscep_exp}. We find that the susceptibility exponents for the symmetric and nematic order parameters change sign at values of $L$ (Fig.~\ref{fig:critical_L})
\begin{align}
L_{c,\text{sym}} ={}& 0.4902 + 0.3490 N,\\
L_{c,\text{nem}} ={}& -7.1975 + 8.3784 N.
\end{align}
Since $L_{c,\text{nem}} \gg L_{c,\text{sym}}$, the nematic criticality seems to be the most fragile, and the first to disappear. The evolution of the correlation length exponent $\nu$ is illustrated at Fig.~\ref{fig:correl_length_exp}.

At first sight it must seem odd that even when $L=1$ we find eight fixed points, at least for $N$ large enough ($N>14.2252$). This is an artifact of the redundancy of our representation of the Gross-Neveu interaction in terms of three seemingly different but ultimately identical quartic terms. All the fixed points in the Gaussian plane are equivalent. This is evident in Fig.~\ref{fig:suscep_exp}, where  the fixed points \subref{fig:suscep_exp_a}, \subref{fig:suscep_exp_c}, \subref{fig:suscep_exp_d}, and \subref{fig:suscep_exp_g} all have the same values of all $x_i =0 $ when $L=1$. Similarly in the ``Gross-Neveu'' plane $g=-1/(4(N-1))$ all the fixed points \subref{fig:suscep_exp_b}, \subref{fig:suscep_exp_e}, \subref{fig:suscep_exp_f}, and \subref{fig:suscep_exp_h} have the exponent for the singlet order parameter $x_i =1 $, and for the other two possible order parameters equal to $x_i = 0 $. When $L>1$, on the other hand, all eight fixed points immediately become distinct.

Returning to the issue of the transition into the symmetric tensor order parameters, we see that the corresponding critical point, at least for $N>14.2252$, requires a value of $L$ of the same order as $N$ or larger for its existence. When $L \rightarrow \infty$ the fixed point corresponds to the mean-field solution of Ref. \cite{so8}. To obtain the true $O(1/L)$-corrections to the mean-field values of the exponents $\gamma=\nu=1$ one also needs the two-loop terms to the beta-functions in Eqs.~\eqref{eq:g1_RG}-\eqref{eq:g3_RG}. Below the critical value $L_{c,\text{sym}} \approx 0.35 N$, for $N\gg1$, however, the symmetric critical point loses its diverging susceptibility, and this way becomes {\it unphysical}.  It would be interesting to address the criticality and the loss thereof at the symmetric-tensor transition within the Gross-Neveu-Yukawa formulation near $3+1$ dimensions \cite{herbutfisherbook, zinnjustin, scherer, zerf, ihrig}.  This work is in progress.

To summarize, we have redefined the canonical Gross-Neveu field theory of interacting $2N$-component Dirac fermions in $2+1$ dimensions in terms of $L$ copies for $4N$-component Majorana fermions. Our extension has three interaction terms which are linearly independent when $L>1$, and it reduces to the standard Gross-Neveu model when $L=1$. For $L\gg N$ the renormalization group transformation, besides the standard Gross-Neveu critical point, displays two new critical fixed points that  describe continuous transitions into mass-order-parameters which transform as the symmetric tensor, and into nematic order parameters that transform as the adjoint (antisymmetric tensor) under $\text{SO}(2N)$. For $N>14.2252$ we find a critical $L_{c} (N)>1$ below which both of these critical points cease to exhibit any diverging susceptibility, and at $L=1$ they become equivalent to the Gaussian fixed point. We interpret this metamorphosis of the symmetric-tensor fixed point in particular as that the spontaneous symmetry breaking $\text{SO}(2N) \rightarrow \text{SO}(N)\times \text{SO}(N)$ in the Gross-Neveu model is likely to be made discontinuous by the quantum fluctuations.

This work was supported by the NSERC of Canada.


\begin{thebibliography}{99}
\bibitem{comment} In contrast to this procedure, introducing $L$ copies of the original $2N$-component Dirac fermions in the Gross-Neveu interaction in Eq.~\eqref{eq:symm_int} would make the theory $\text{SO}(2LN)$-symmetric, and would merely amount to rescaling $N\rightarrow L N$.
\end{thebibliography}
\end{document}